\documentclass[10pt]{article} %
\usepackage[preprint]{rlc}

\usepackage{amssymb}            %
\usepackage{mathtools}          %
\usepackage{mathrsfs}           %
\mathtoolsset{showonlyrefs}     %
\usepackage{graphicx}           %
\usepackage{subcaption}         %
\usepackage[space]{grffile}     %
\usepackage{url}                %

\usepackage[linesnumbered,lined,ruled,vlined]{algorithm2e}
\usepackage{amsmath}
\usepackage{amsfonts}
\usepackage{booktabs}
\usepackage{dsfont}
\usepackage{subcaption}
\usepackage{wrapfig}
\usepackage{pgf}
\usepackage[capitalise,noabbrev,nameinlink]{cleveref}

\newcommand{\set}[1]{\mathcal{{#1}}}

\DeclareMathOperator{\nbr}{\mathrm{ne}}

\newcommand\showvictor{0}

\newcommand{\vad}[1]{%
    \ifnum 1=\showvictor \relax
        {\color{cyan}#1}\else #1%
    \fi
}

\title{Trust-based Consensus in\\Multi-Agent Reinforcement Learning Systems}

\author{Ho Long Fung\textsuperscript{\rm 1}, Victor-Alexandru Darvariu\textsuperscript{\rm 1}, Stephen Hailes\textsuperscript{\rm 1}, Mirco Musolesi\textsuperscript{\rm 1,2}\\
    \texttt{\{ho.fung.20, v.darvariu, s.hailes, m.musolesi\}@ucl.ac.uk} \\
    \textsuperscript{\rm 1}University College London \textsuperscript{\rm 2}University of Bologna
    }

\begin{document}

\maketitle

\begin{abstract}
    An often neglected issue in multi-agent reinforcement learning (MARL) is the potential presence of unreliable agents in the environment whose deviations from expected behavior can prevent a system from accomplishing its intended tasks.
    In particular, consensus is a fundamental underpinning problem of cooperative distributed multi-agent systems.
    Consensus requires different agents, situated in a decentralized communication network, to reach an agreement out of a set of initial proposals that they put forward.
    Learning-based agents should adopt a protocol that allows them to reach consensus despite having one or more unreliable agents in the system.
    This paper investigates the problem of unreliable agents in MARL, considering consensus as a case study.
    Echoing established results in the distributed systems literature, our experiments show that even a moderate fraction of such agents can greatly impact the ability of reaching consensus in a networked environment.
    We propose Reinforcement Learning-based Trusted Consensus (RLTC), a decentralized trust mechanism, in which agents can independently decide which neighbors to communicate with.
    We empirically demonstrate that our trust mechanism is able to handle unreliable agents effectively, as evidenced by higher consensus success rates.
\end{abstract}

\section{Introduction}

A \emph{cooperative multi-agent system} (MAS) is a system composed of multiple autonomous entities, known as \emph{agents},  which collaborate within a shared environment to solve tasks in order to improve their joint welfare \citep{Dafoe2020OpenCoop}. A largely neglected yet fundamental issue in the multi-agent reinforcement learning (MARL) literature is the potential presence of \textit{unreliable} agents within the environment. Indeed, dealing with unreliability is of essential importance for building real world  MARL systems. An agent might be unreliable due to node and/or transmission failure, or simply because it exchanges incorrect information. Identifying the exact cause of unreliability is difficult and often impossible a priori.

A general class of cooperative tasks that is susceptible to unreliable agents is \emph{consensus}.
Consensus is indeed an essential element of the design of any distributed system, including multi-agent systems. It is one of the classic problems in multi-agent and distributed systems, which has fascinated generations of scientists from the foundations of the field \citep{LamportByzantine,Lamport1998Paxos} due to its theoretical elegance and vast applicability to a series of practical problems \citep{Ongaro2014Raft,Olfati2007NetworkedConsensus,Barrat2009ComplexNetworks}.
Consensus deals with reaching an agreement among agents that put forward different proposals. A \emph{consensus protocol} can be defined as a procedure that agents follow in order to reach agreement successfully.
A protocol can be hard-coded,
where each agent executes an algorithm written by an expert
(e.g., in the distributed systems literature \citep{Cachin2011Reliable}).
Otherwise, the protocol is \emph{emergent} if it arises
from agents learning how to achieve consensus through repeated interactions.
Consensus problems are often studied in decentralized settings,
where agents can only make local observations of the environment,
act independently of each other
and interact solely by communicating over a network
\citep{LamportByzantine,PSL1980AgreeFaults,Cachin2011Reliable,Coulouris2012}.
These limitations make the problem of reaching consensus nontrivial.
A reason is the lack of a central coordinator that can aggregate and distribute values to and from all agents in the system.
Another reason is the existence of failed or unreliable agents in the network
that deviate from their expected behavior.
Decentralization renders failures hard to detect since this cannot be performed based on direct observation.

In real-world scenarios,
the occurrence of unexpected failures is often an integral property of the environment~\citep{Schroeder2010FailHPC,Cloudflare2020Byzantine}.
Suppose that a MAS adopts a consensus protocol that assumes agents are always reliable.
In cases where agents fail unexpectedly, this may lead to errors in message transmission or performing local computations.
Since the protocol cannot accommodate for these failures,
agents will assume that the rest of the system is still reliable and
execute the protocol as usual, using incomplete or incorrect information,
which may severely impact the system's performance.
Therefore, to reach consensus reliably,
agents must adopt a protocol that can continue operating at an acceptable level despite one or more failures in the system.
One way to deal with unreliable agents is through the introduction of \emph{trust mechanisms},
which have been widely studied in the distributed systems
and multi-agent systems literature (e.g.,~\citep{Rahman1998Trust,Ramchurn2004Trust}).
Agents can try to quantify the trustworthiness of other agents
based on past interactions and data about them.

This paper investigates the problem of unreliable agents in MARL, considering the problem of distributed consensus as a case study.
We start by assuming a network of reliable agents with a basic communication model,
then introduce unreliable agents and observation noise and study their impact on achieving consensus.
In line with classic results in distributed systems (e.g., ~\cite{LamportByzantine,PSL1980AgreeFaults}), we find that unreliable agents greatly impact the ability of decentralized learning-based setups to reach consensus. To mitigate this issue, we introduce the notion of trust by equipping RL agents with a learnable trust mechanism,
which we refer to as \emph{Reinforcement Learning-based Trusted Consensus (RLTC)}.
Experiments show that RLTC increases the consensus success rate of the system
when compared to a setting without the trust mechanism
(i.e., in which agents assume all other agents are reliable),
generalizing to various types of failure models. 
We also show that RLTC is able to scale as the number of agents increases, demonstrating practical applicability of the proposed mechanism to real-world scenarios.
\section{Related Work}

\noindent \textbf{Emergent communication protocols in MARL.}
In the last decade, there has been a significant amount of work on deep multi-agent reinforcement learning for cooperative multi-agent tasks,
such as traffic junctions \citep{Sukhbaatar2016}, multi-robot warehouses \citep{Christianos2020shared}, cooperative navigation and predator-prey games \citep{Lowe2017multi}.
Differentiable communication mechanisms have been introduced to allow deep agents to choose what and to whom to communicate \citep{Foerster2016DIAL,Sukhbaatar2016,Das2019Tarmac,RangwalaWilliams2020,Zhang2020I2C}.
Progress is also being made in the design of \emph{emergent communication protocols},
where agents learn to associate communication symbols
with perceptual input and actions by solving cooperative downstream tasks
\citep{Mordatch2017,Lazaridou2018,Bouchacourt2019,Graesser2019}.
However, 
this body of work often assumes that all agents are fully functional throughout the lifetime of the system.
In our work, we instead investigate if RL agents can learn to adapt to the presence of unreliable agents.

\vad {
    \noindent \textbf{Adversarial attacks in (MA)RL.} Several works have studied the design of RL policies that are robust to adversarial attacks in a single-agent setting.~\citet{pinto2017robust} focused on continuous control environments in which an adversary is allowed to apply disturbances to the agent's action, demonstrating that more robust policies can be obtained even in the absence of the adversary at test time.~\citet{zhang2020robust} considered a threat model in which the observations received by the agent are adversarially perturbed. In the MARL literature,~\citet{blumenkamp2021emergence} showed that, in environments that are not fully cooperative, self-interested agents are able to learn to construct messages that disrupt the cooperative agents.~\citet{xue2022mis} proposed to train a model that recovers the true message from the malicious message, which relies on the assumption that the message is perturbed (rather than replaced altogether).~\citet{sun2023certifiably}  considered a multi-agent setting and proposed a technique for dealing with adversarial agents that relies on aggregating the received messages (by majority vote for the discrete action case and the median for each coordinate in the continuous case). This technique requires the assumption of a non-adversarial training phase for learning the message aggregation policy. In contrast, RLTC is based on an explicit and interpretable trust mechanism that improves performance over standard MARL even when unreliable agents are present at training time. 
}

\noindent \textbf{Consensus in complex networks.}
Interactive models have been used to study the
collective behavior of agents in networked environments,
such as social networks and particle systems in general.
Examples include the Voter model
\citep{Barrat2009ComplexNetworks} and
majority rule \citep{Krapivsky2003Majority},
where neighboring agents can communicate and update their local state using a random mechanism.
In our work, we use a discrete-time variant of the Voter model,
where each agent can choose who to communicate with
by sampling from a subset of trusted neighbors, which the agent can adjust over time.

\noindent  \textbf{Averaging consensus algorithms.}
There are other formulations of consensus applied to areas such as social influence networks \citep{DeGroot1974Social}, sensor networks \citep{Olfati2005Sensors} and vehicle formations \citep{Fax2004Vehicles}.
A relevant example is the class of \emph{averaging algorithms},
where each node in a network updates its local scalar value by computing an average over its neighbors, potentially using different and/or randomized weights \citep{Bullo2022Networks}.
Properties of these algorithms, such as asymptotic convergence, have been studied with matrix theory and algebraic graph theory \citep{Ren2005Survey,Olfati2007NetworkedConsensus}.
In contrast, our work uses RL, where each agent decides which neighbors to trust and aggregate values from, resulting in adaptive rather than pre-determined behavior.

\noindent  \textbf{Trust in distributed systems.}
\label{par:related-trust}
Trust is important for supporting cooperation and coordination
in multi-agent systems \citep{Dafoe2020OpenCoop}.
Various formulations of trust have been studied in the distributed and multi-agent systems literature.
Ways to quantify trust include discrete trust levels and scores, which are usually computed based on past interactions and data about other agents \citep{Rahman1998Trust,Sabater2005TrustReview}.
These can help an agent to decide if it is beneficial to cooperate with another agent.
Models based on reputation and recommendations for gathering, aggregating and promoting ratings
that approximate the trustworthiness of an agent have also been developed~\citep{Rahman1998Trust,Castelfranchi1998Trust,Ramchurn2004Trust,Sabater2005TrustReview,Cohen2019Trust}.
Our work is related to the area of decentralized trust,
where agents learn who to trust by interacting with others,
rather than relying on a central authority \citep{Rahman1998Trust}.

\section{Problem Description}
This section provides a formal definition of a consensus problem,
including modeling assumptions such as the means of inter-agent communication and trust mechanisms.

\noindent\textbf{Consensus definition.}
\label{par:consensus}
Consensus is reached when agents agree upon a common value out of a set of initial proposals.
Denote the set of agents in the system as $\set{N} = \{1, 2, \dots, N\}$.
They must agree upon a global value from a known set of candidates $\set{D}$,
which we assume is binary: $\set{D} = \{0, 1\}$.
We also define a predetermined ground-truth value as 1,
while the incorrect value is 0.
Initially, each agent $i \in \set{N}$ proposes a value $v_i \in \set{D}$ with independent probability $p$
of being equal to 1 and 0 with probability $1 - p$,
the latter of which we will refer to as the \emph{noise parameter}.
Then, agents can share information over a communication network and update their local values appropriately.
Eventually, consensus is reached if all agents agree upon the true value, otherwise known as \emph{global consensus}.
However, due to the decentralized nature of our setup,
where agents can only interact with their immediate neighbors, it is difficult for agents to determine if all other agents agreed on 1.
Thus, we define an agent-specific success criterion,
known as \emph{local consensus}, which is when the agent and its neighbors in the network agree on 1.

\begin{wrapfigure}{r}{0.22\linewidth}
  \vspace{-0.35\baselineskip}
  \begin{center}
    \includegraphics[width=\linewidth]{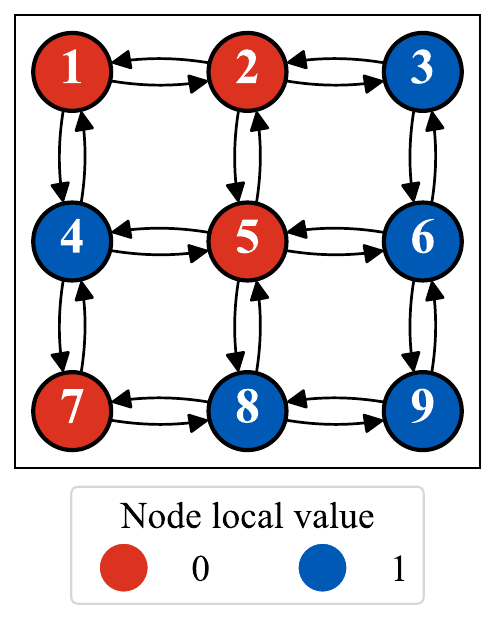}
    \caption{
        Communication network example with 9 agents, where nodes represent agents, arrows indicate communication links and colors are mapped to the agents' local values.
    }
    \label{fig:communication-9}
  \end{center}
  \vspace{-\baselineskip}
\end{wrapfigure}

\begin{figure}[t]
  \centering
  \includegraphics[width=0.70\textwidth]{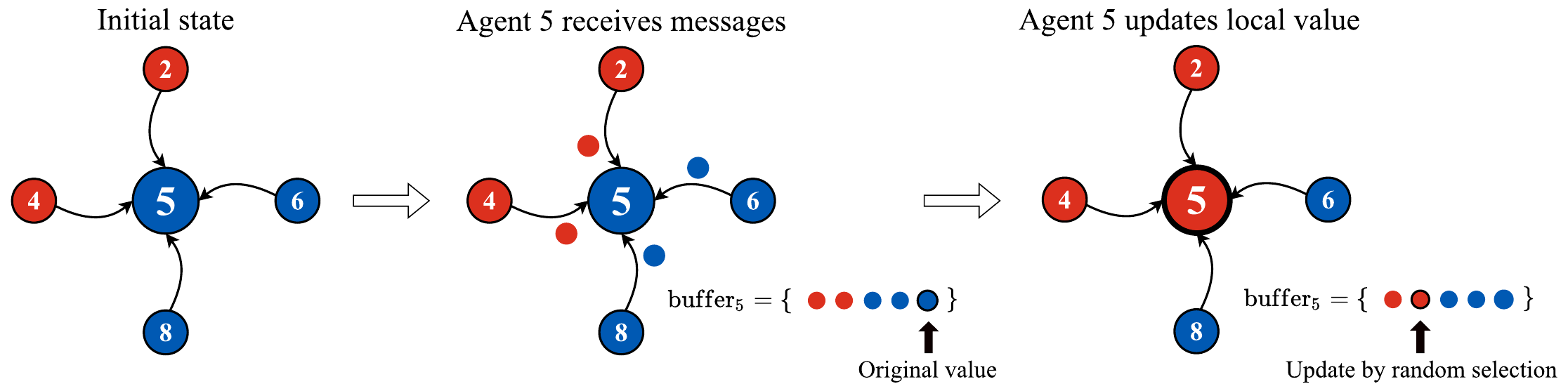}
  \caption{
      Communication mechanism from the perspective of agent 5.
      Agent 5 receives the values from its neighbors,
      then updates its local value by randomly selecting from the
      set of received values and its own.
      During each timestep, this is performed simultaneously by all agents.
  }
  \label{fig:communication-one-agent-full-trust}
\end{figure}

\noindent\textbf{Communication model.}
\label{par:communication}
Agents are interconnected by a communication network.
It is defined by an undirected, connected graph $\set{G} = (\set{N}, \set{E})$
with nodes $\set{N}$, which correspond to agents, and bi-directional communication links $\set{E}$ between nodes.
For simplicity, we assume the communication network is a 2D square lattice.
See Figure \ref{fig:communication-9} for an illustration with $N = 9$ agents.
Agents interact by performing consecutive message-passing rounds in lock-step.
A round starts with each agent broadcasting its current value $v_i$ to all of its neighbors $\mathrm{ne}(i)$.
Then, upon receiving all incoming values into a message buffer denoted $\mathrm{buffer}_i$, each agent updates its current value by randomly selecting a value in $\{v_i\} \cup \mathrm{buffer}_i$.
See Figure \ref{fig:communication-one-agent-full-trust} as an example of communication between an agent and its neighbors.
This communication mechanism described is similar to opinion formation models
which are used to simulate interactions in social networks
\citep{Barrat2009ComplexNetworks}.
They assume a population of agents that initially have contradictory opinions
and can update their own opinions over time by interacting pairwise with other agents in the network.
This is also similar to randomized averaging algorithms, where nodes can choose which values to aggregate by following a stochastic model \citep{Bullo2022Networks}.

\begin{wrapfigure}{r}{0.61\linewidth}
  \vspace{-1.25\baselineskip}
    \centering
    \includegraphics[width=\linewidth]{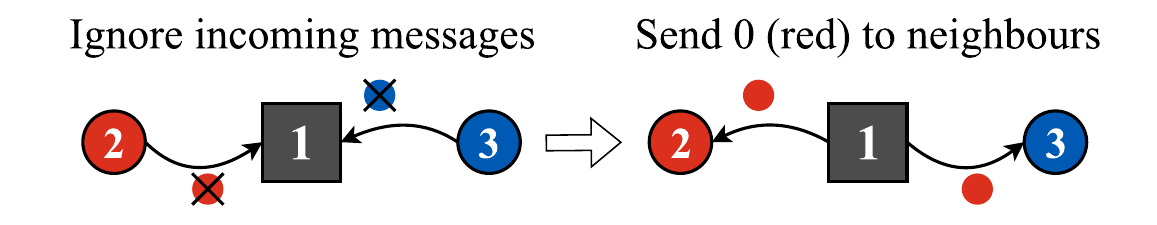}
    \caption{Example of an unreliable agent 1 (square) that always ignores messages from neighbors 2 and 3 and sends 0.}
    \label{fig:communication_one_agent_unreliable}
  \vspace{-\baselineskip}
\end{wrapfigure}

\noindent\textbf{Failure models.}
\label{par:failure-model}
In addition, we specify the \emph{failure models} that describe how agents can deviate from expected behavior.
Here, we assume that a faulty or unreliable agent may send the incorrect value 0 to its neighbors.
This is shown in Figure \ref{fig:communication_one_agent_unreliable}.
If an unreliable agent does this all the time, we refer to this as following a \emph{Fixed} failure model.
Another possibility is that the agent sends 0 or 1 uniformly at random,
which we refer to as the \emph{Random} failure model.
We will investigate the extent of which our trust-based mechanism that we introduce next can generalize to both types of unreliability.
In addition, Figure \ref{fig:example-full} illustrates an example network with a single unreliable Fixed agent in which, despite the simple failure model, consensus is not achieved due to the propagation of misinformation by the other agents.
Note that agents cannot determine the reliability of other agents a priori due to the decentralized property of the system, hence learning is necessary.

\section{Trust Mechanisms for Consensus}
\label{section:trust-mechanism}

A potential solution to our consensus problem is through the notion of trust, as introduced in Section \ref{par:related-trust}.
To this end, we propose a simple decentralized trust mechanism that can be learned through MARL.
In this section, we present its basic operation, then Section \ref{section:marl-model} will provide the reader with a more formal description of the proposed solution.

\begin{figure}[h]
  \centering
  \includegraphics[width=0.70\textwidth]{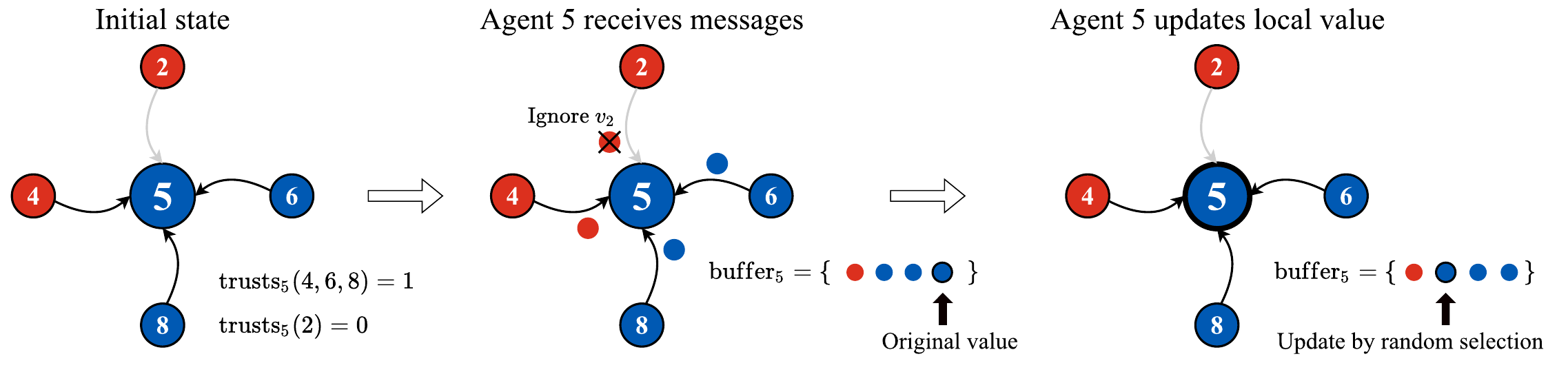}
  \caption{
      Communication between agent 5 and its neighbors, but agent 5 is equipped with a trust mechanism.
      Agent 5 trusts neighbors 4, 6 and 8 but not agent 2.
      It updates its local value by randomly sampling
      from itself and its \emph{trusted} neighbors only.
  }
  \label{fig:communication-one-agent-partial-trust}
\end{figure}

Each agent $i$ maintains a binary score for each neighbor $j$, representing whether $i$ trusts $j$.
Denote $\mathrm{trusts}_i(j)$ as the trust score of agent $i$ towards
$j$ and $\mathrm{trusts}_i(\cdot)$ as the array containing all trust scores of that agent.
If an agent $i$ distrusts a neighbor $j$,
or $\mathrm{trusts}_i(j) = 0$, then it can choose to ignore all incoming messages from $j$.

\begin{wrapfigure}{r}{0.28\linewidth}
  \vspace{-1.2\baselineskip}
  \centering
  \includegraphics[width=0.99\linewidth]{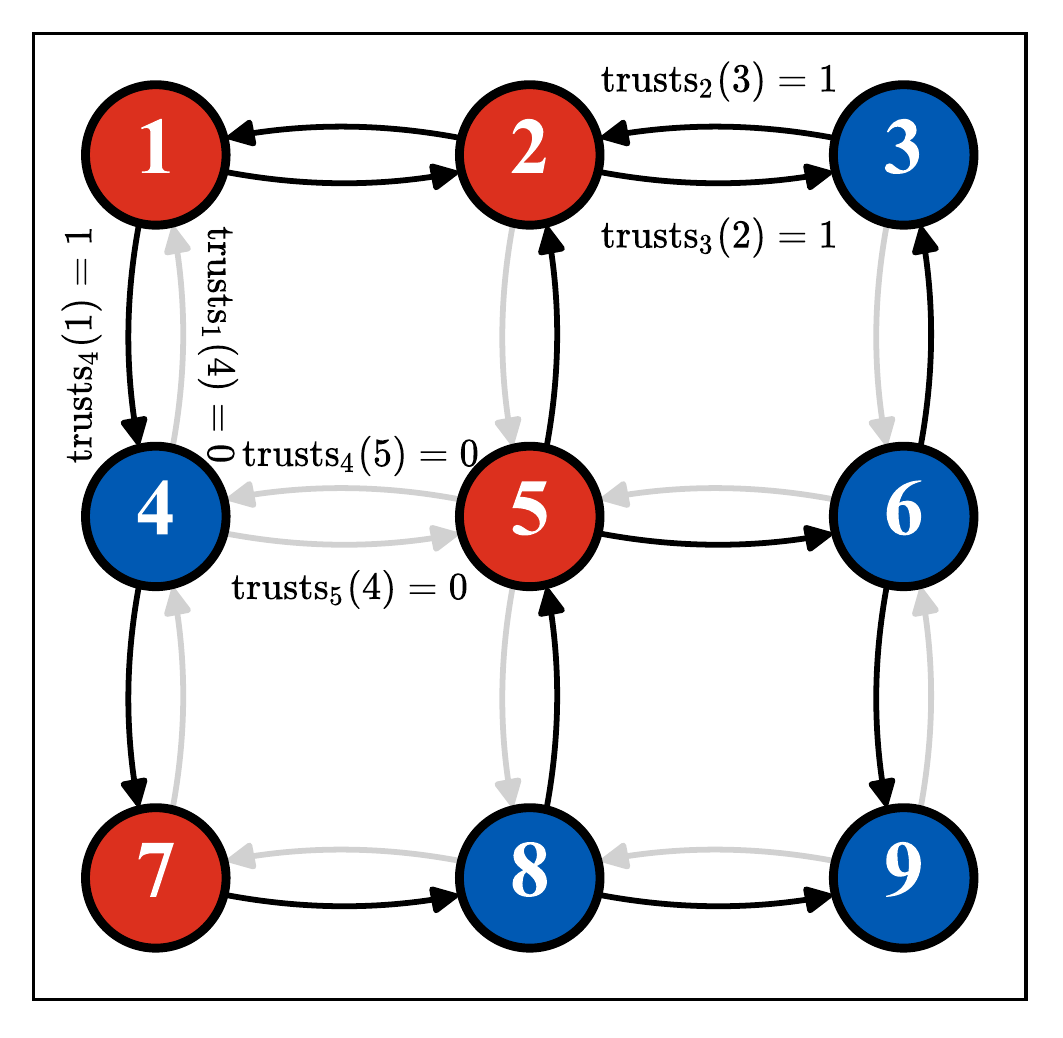}
  \caption{
      Communication grid example with 9 agents, in which not all agents trust each other.
  }
  \label{fig:communication-trust-score}
  \vspace{-\baselineskip}
\end{wrapfigure}

In effect, the array $\mathrm{trusts}_i(\cdot)$ defines the subset of neighbors from which agent $i$ is allowed to sample values from during each timestep.
Figure \ref{fig:communication-one-agent-partial-trust} demonstrates this from the perspective of one agent.
Figure \ref{fig:communication-trust-score} illustrates an example of trust in a grid.
For example, both agents 4 and 5 distrust each other
($\mathrm{trusts}_4(5) = 0$ and $\mathrm{trusts}_5(4) = 0$).
As a result, both agents ignore each other's incoming messages, as indicated by the grayed out arrows from node 5 to 4 and vice versa.
Agents 2 and 3 mutually trust each other, as shown by black arrows in both directions.
Agent 1 distrusts Agent 4 (so Agent 1 ignores 4's messages, indicated by the grayed out \emph{incoming} arrow into Agent 1 from Agent 4), but Agent 4 trusts Agent 1. This is an example of unreciprocated trust.
Through continual interactions, each agent can update its trust scores
in order to minimize the effects of unreliable agents and improve the chances of successful consensus.

\SetKwComment{Comment}{/* }{ */}
\begin{algorithm}[t]
\begin{footnotesize}
  \caption{RLTC Consensus Protocol Episode Execution}
  \label{algorithm:consensus}
  Initialize each $v_i$ randomly to 1 with probability $p$, 0 otherwise. \\
  Initialize all $\mathrm{trusts}_i(j) = 1$. \\
  \For{timestep $t = 1, 2, \dots, T$}{
      \For{$i \in \mathcal{N}$}{
        \tcp{Receive messages}
          $\mathrm{buffer}_i \gets \{ v_j \ \vert \ j \in \nbr(i) \land \mathrm{trusts}_i(j) = 1 \}$
      }
      \For{$i \in \mathcal{N}$}{
          \tcp{Update local values}
          $v_i \gets \mathrm{random}(\{v_i\} \cup \mathrm{buffer}_i)$
      }
      \For{$i \in \mathcal{N}$}{
          $j \gets \pi^i(s^i_t)$; \\
          \If{$j \neq \emptyset$}{
            \tcp{Update trust score}
            $\mathrm{trusts}_i(j) \gets \neg \mathrm{trusts}_i(j)$
          }
          \tcp{
              If training, do $Q$-learning update for agent $i$  (Eq. \ref{eq:q-learning-indep})
          }
      }
  }
  \end{footnotesize}
\end{algorithm}

\begin{table*}[h!]
\begin{footnotesize}
  \centering
  \begin{tabular}{p{0.22\linewidth} p{0.43\linewidth} p{0.35\linewidth}}
      \toprule
      Name & Equation & Description \\
      \midrule
      Success rate &
      \(\displaystyle
      \frac{1}{N} \sum_{i=1}^N \mathds{1}[v_i = v_\mathrm{true}] \) &
      Fraction of reliable agents with the correct value.\\
      Average trust rate &
      \(\displaystyle
      \frac{1}{N}
      \sum_{i \in \mathcal{N}}
      \frac{1}{\lvert \mathrm{ne}(i) \rvert}
      \sum_{j \in \mathrm{ne}(i)} \mathrm{trusts}_i(j) \) &
      Average fraction of neighbors, reliable or otherwise, that each reliable agent trusts. \\
      Mutual trust rate &
      \(\displaystyle
      \frac{1}{\lvert \mathcal{E} \rvert}
      \sum\limits_{i, j \in \mathcal{E}}
          \mathds{1}[ \mathrm{trusts}_i(j) \land \mathrm{trusts}_j(i) ]
      \) &
      Fraction of mutually-trusting reliable agents.
      0 if no reliable agents are adjacent in the network.\\
      Average trust accuracy &
      \(\displaystyle
      \frac{1}{N}
      \sum\limits_{i \in \mathcal{N}} \frac{1}{\lvert \mathrm{ne}(i) \rvert}
          \sum\limits_{j \in \mathrm{ne}(i)} \mathds{1}[
              \mathrm{trusts}_i(j) = \mathds{1}[j \in \mathcal{N}]]
      \) &
      Fraction of correct trust scores assigned to neighbors on average,
      i.e., the agent assigns trust score 1 for reliable neighbors and 0 for unreliable ones. \\
      \bottomrule
  \end{tabular}
\caption{Performance metrics used in the evaluation.}
\label{table:metrics}
\end{footnotesize}
\end{table*}

\section{MARL Model}
\label{section:marl-model}
\begin{figure}[h!]
  \centering
  \begin{subfigure}{0.8\linewidth}
      \centering
      \includegraphics[width=\linewidth]{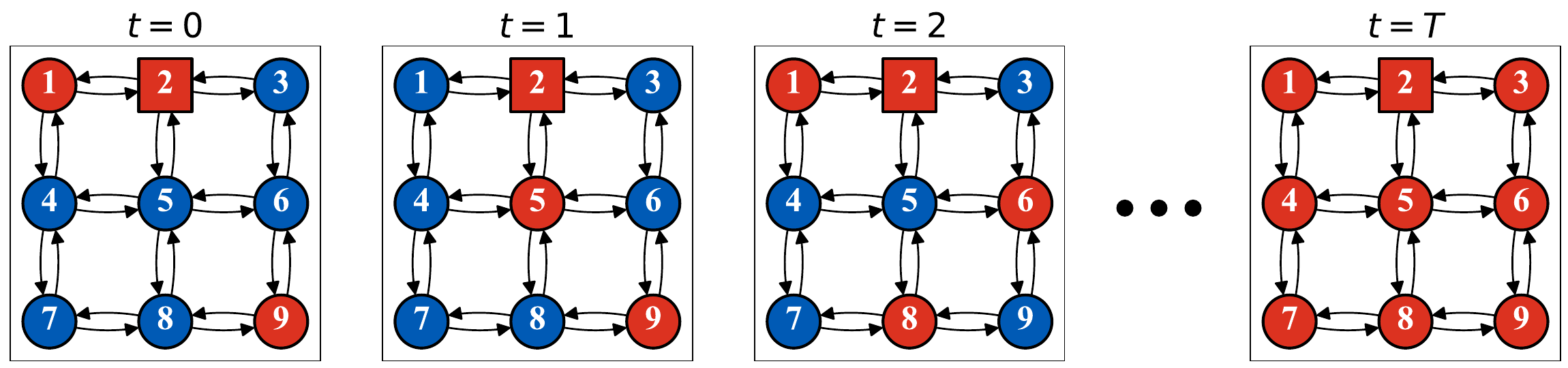}
      \caption{Without trust mechanism (implicit full trust).}
      \label{fig:example-full}
  \end{subfigure}
  \begin{subfigure}{0.8\linewidth}
      \centering
      \includegraphics[width=\linewidth]{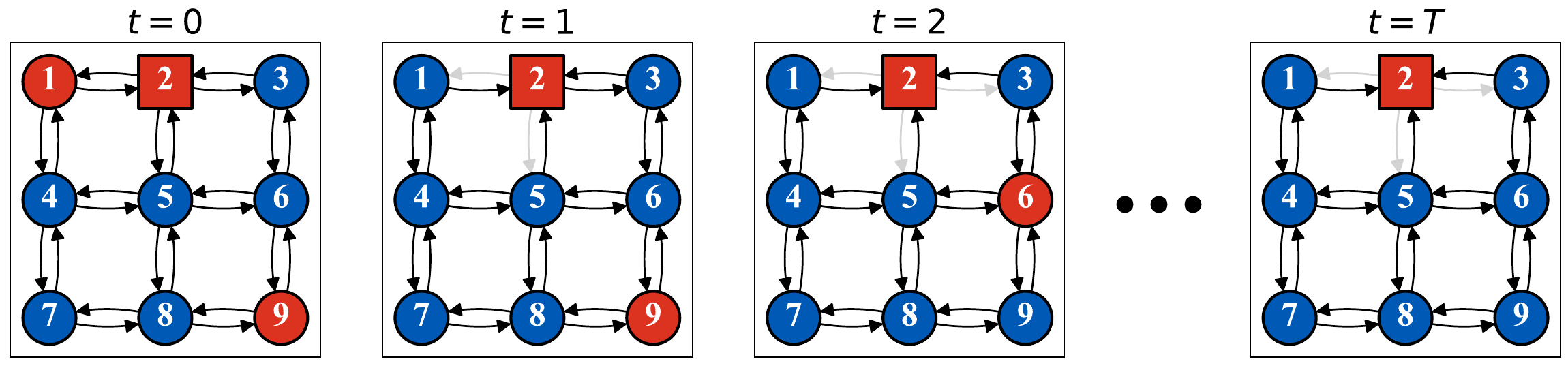}
      \caption{With trust mechanism (agents 1, 5 and 3 distrust agent 2).}
      \label{fig:example-trust}
  \end{subfigure}
  \caption{
    An example scenario with 9 agents with one unreliable agent labeled 2.
    Timesteps 0, 1, 2 and $T = 30$ of an episode are shown.
    Figure \ref{fig:example-full} does not have the trust mechanism and results in failure, because the incorrect value 0 (red) has spread to all agents.
    Figure \ref{fig:example-trust} has the trust update mechanism.
    Reliable agents 1 and 5 toggle their trust scores of agent 2 to 0 at $t = 1$, then agent 3 does the same at $t = 2$.
    Here, consensus is successful, because agent 2 is effectively isolated from the rest of the network, and all reliable agents agree on 1 (blue).
  }
  \label{fig:episode-example}
\end{figure}

Given the trust mechanism discussed in Section \ref{section:trust-mechanism},
we will describe how the agents can learn to update their trust scores.
This is achieved by formulating the problem
as a multi-agent decentralized Markov Decision Process, then learning the trust mechanism using $Q$-learning \citep{Watkins1992,Tan1993IndepQ}.
The multi-agent MDP that we will use is characterized by a tuple
$\set{M} = (
  \set{S},
  \set{A},
  P,
  R
)$,
containing the state space, action space, transition probability and reward function respectively.
Each of these elements is described below.
In addition, we assume a finite horizon of length $T$.
Note that each agent $i$ in the MDP corresponds to a \emph{reliable} node in the network only.
Unreliable nodes follow a pre-defined behavior as described in Section \ref{par:failure-model}.
They do not have a trust mechanism and are not trained, thus are excluded from the MDP.
Figure \ref{fig:example-trust} illustrates an example of episodic execution with 9 agents.

\noindent \textbf{States.}
$\set{S} = \prod_{i \in \set{N}} \set{S}^i$ is the global state space, which is composed of the local state spaces $\set{S}^i$ per agent $i$.
An agent's local state $s^i$ corresponds to its trust array $\mathrm{trusts}_i(\cdot)$
as described in Section \ref{section:trust-mechanism}.
In other words, $s^i$ maps neighbors to trust scores: $s^i : \mathrm{ne}(i) \mapsto \{0, 1\}$.
For example, Agent 1 in Figure \ref{fig:communication-trust-score} has
$\mathrm{trusts}_1(2) = 1$ and $\mathrm{trusts}_1(4) = 0$,
so $s^1 = \{(2, 1), (4, 0)\}$.
Agent 5's local state is $s^5 = \{(2, 0), (4, 0), (6, 0), (8, 1)\}$.
At the start of each episode, all entries of all trust arrays are initialized to 1
(e.g., $s^5 = \{(2, 1), (4, 1), (6, 1), (8, 1)\}$ at $t=0$).
Note that local states only consider trust scores and are independent of the agent's current local value (0 or 1).

\noindent \textbf{Actions.}
$\set{A} = \prod_{i \in \set{N}} \set{A}^i$ is the joint action space composed of the individual local action space $\set{A}^i$ of each agent $i$.
A local action $a^i$ corresponds to agent $i$ toggling its trust score of a specific neighbor or not toggling any score at all.
So, $\set{A}^i = \{\emptyset\} \cup \mathrm{ne}(i)$, where $\emptyset$ is a no-op action.
Concretely, if $a^i = j$ then $\mathrm{trusts}_i(j) \gets \neg \mathrm{trusts}_i(j)$.
For example, in Figure \ref{fig:example-trust} between $t=0$ and $t=1$, agent 1 toggles its trust score of neighbor 2 from $\mathrm{trusts}_1(2) = 1$ to 0 by selecting action $a^1 = 2$.
Agent 4 did not modify its trust scores, so $a^4 = \emptyset$.
Due to the decentralized nature of our setup, agents select actions $a^i$ independently, following their local policy $\pi^i(s^i)$ and independently of other agents.

\noindent \textbf{Transitions.}
$P$ is the transition function:
$P(s' \vert s, \{a^i\}_{i \in \set{N}})$
is the probability of moving from global state $s \in \set{S}$ to state $s' \in \set{S}$ when each agent $i$ executes some action $a^i$.
The global state is the union of the trust score arrays of all agents in the system
after they individually perform local actions (toggling or doing nothing).
The state transitions are deterministic, only depending on the agents' actions.
See Figure \ref{fig:example-trust} for examples of consecutive global state transitions,
but ignore the agents' local values (red or blue) because the MDP does not consider this information.

\noindent   \textbf{Reward and consensus optimization objective.}
The goal of the trust mechanism is to ensure that consensus can still be attained despite disruptions caused by unreliable nodes in the network.
Each reliable agent must update its trust scores so that it reaches agreement with its neighbors correctly.
To achieve this, it must discover a policy $\pi^i$ that allows it to perform appropriate updates to its trust array.
Thus, we define the reward function $R^i$ in terms of the local success criteria for consensus described in Section \ref{par:consensus}, which is appropriate for our decentralized setup where agents can only interact directly with their neighbors.
Each agent $i$ is assigned reward $R^i = +1$ if it correctly agrees with all neighbors:
$\forall j \in \{i\} \cup \nbr(i): \ v_j = 1$.
If this is not satisfied, then $R^i = -1$.
The global objective is to find an optimal joint policy
$\pi = \langle \pi^1, \dots, \pi^N \rangle$
that maximizes the joint action-value function
$Q^\pi(s, a) = \mathbb{E}\left[ \sum_{k=t+1}^{T} \gamma^{k - t - 1} \sum_{i \in \set{N}} R_k^i \vert s_t = s, a_t = a\right]$,
where $s$ is the joint state,
$a$ is the joint action,
$T$ is the horizon and $\gamma \in [0, 1)$ is the discount factor.

\noindent \textbf{$Q$-learning update.}
Due to the decentralized setup,
we will optimize the above objective using
independent $Q$-learning \citep{Tan1993IndepQ}, where each agent learns according to its own actions and local state.
The update rule for agent $i \in \{1, \dots N\}$ is as follows:
\begin{equation}
  \label{eq:q-learning-indep}
  \begin{split}
  Q^i(s_t^i, a_t^i)
    \gets Q^i(s_t^i, a_t^i) + \alpha \left(
    \left[R^i_{t} + \gamma \max_{a \in \set{A}^i} Q^i(s_{t+1}^i, a)\right]
    - Q^i(s_t^i, a_t^i)
  \right)
  \end{split}
\end{equation}
where $Q^i$ is the $Q$-function of agent $i$, $\alpha$ is the learning rate and $\gamma$ is the discount factor.
The update uses states and actions local to agent $i$, i.e., $s^i$ and $a^i$.
Agents use the $\epsilon$-greedy policy for training and the greedy policy for evaluation.

\noindent \textbf{Summary of consensus protocol execution.}
The full execution of our consensus protocol is detailed in Algorithm \ref{algorithm:consensus}. To aid understanding, an example is illustrated in Figure \ref{fig:example-trust}.
At the beginning, the communication graph contains both reliable and unreliable agents.
In the example, only agent 2 is unreliable.
Reliable agents are trained, and unreliable agents have fixed behavior as described in Section \ref{par:failure-model}.
Then, the consensus protocol consists of both the trust mechanism (set up as an MDP above)
and communication (sending and updating local values, which are not part of the MDP).
The protocol executes over a finite episode with length $T$.
At the start of each episode,
all trust scores are initialized to 1,
as indicated by all black arrows in Figure \ref{fig:example-trust}.
Each reliable agent's local value is randomly set to 0 with probability $1-p$, 1 otherwise,
while the unreliable agents always start with 0.
In the example, reliable agents 1 and 9 start with 0 while the others have 1,
while unreliable agent 2 has 0.
Then, each timestep consists of three phases:
agents receive the incoming values of trusted neighbors, update their local values, and finally update their trust scores (decided by the agents' local action policies).
The episode terminates after the final timestep $T$ is reached.

\begin{figure}[t]
    \centering
    \includegraphics[width=0.85\linewidth]{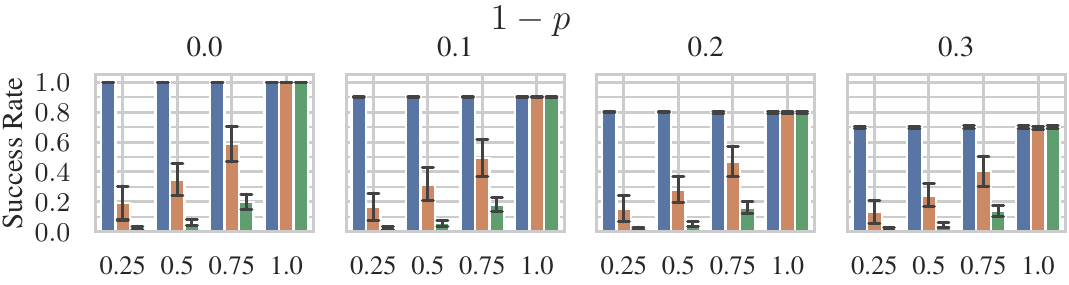} \\
    \includegraphics[width=0.85\linewidth]{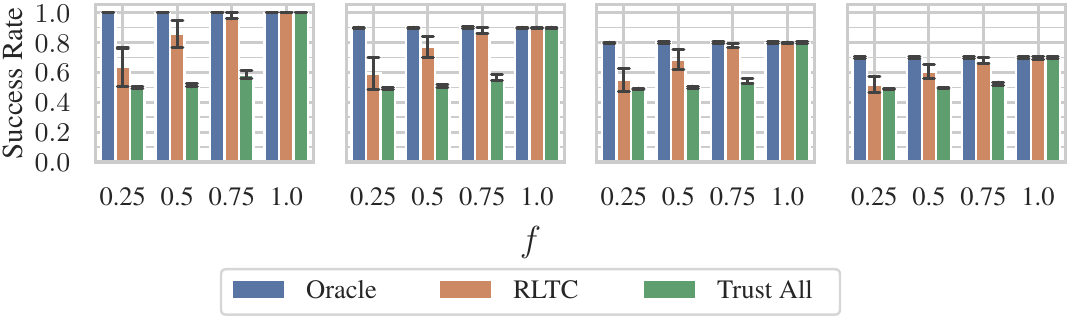}
    \caption{
        Success rates for different fractions of reliable agents $f$ and noise values $1-p$ (16 agents).
        Each bar height and error bar represent the mean
        and 1 standard deviation respectively wrt. 30 runs.
        Top and bottom plots are for the \emph{Fixed} and \emph{Random} failure models respectively.
    }
    \label{fig:success-rates}
\end{figure}
\begin{figure}[t]
    \begin{center}
        \includegraphics[width=0.89\linewidth]{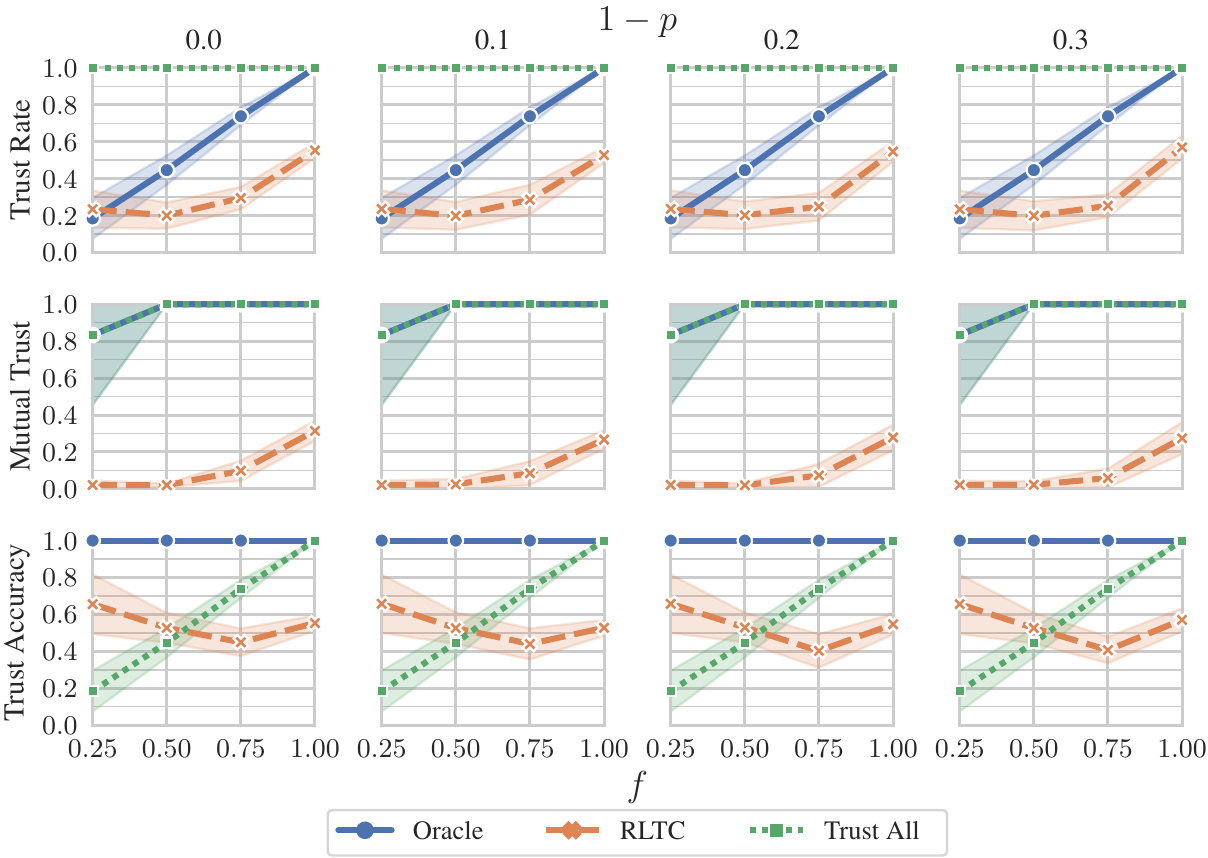}
    \end{center}
    \caption{
        $f$ against trust metrics for fixed $1-p$ with 16 agents and the \emph{Fixed} failure model.
        Each line and error band represent the mean and 1 standard deviation respectively wrt. 30 runs.
    }
    \label{fig:frac-reliable}
\end{figure}
\begin{figure}[t]
    \begin{center}
        \includegraphics[width=0.89\linewidth]{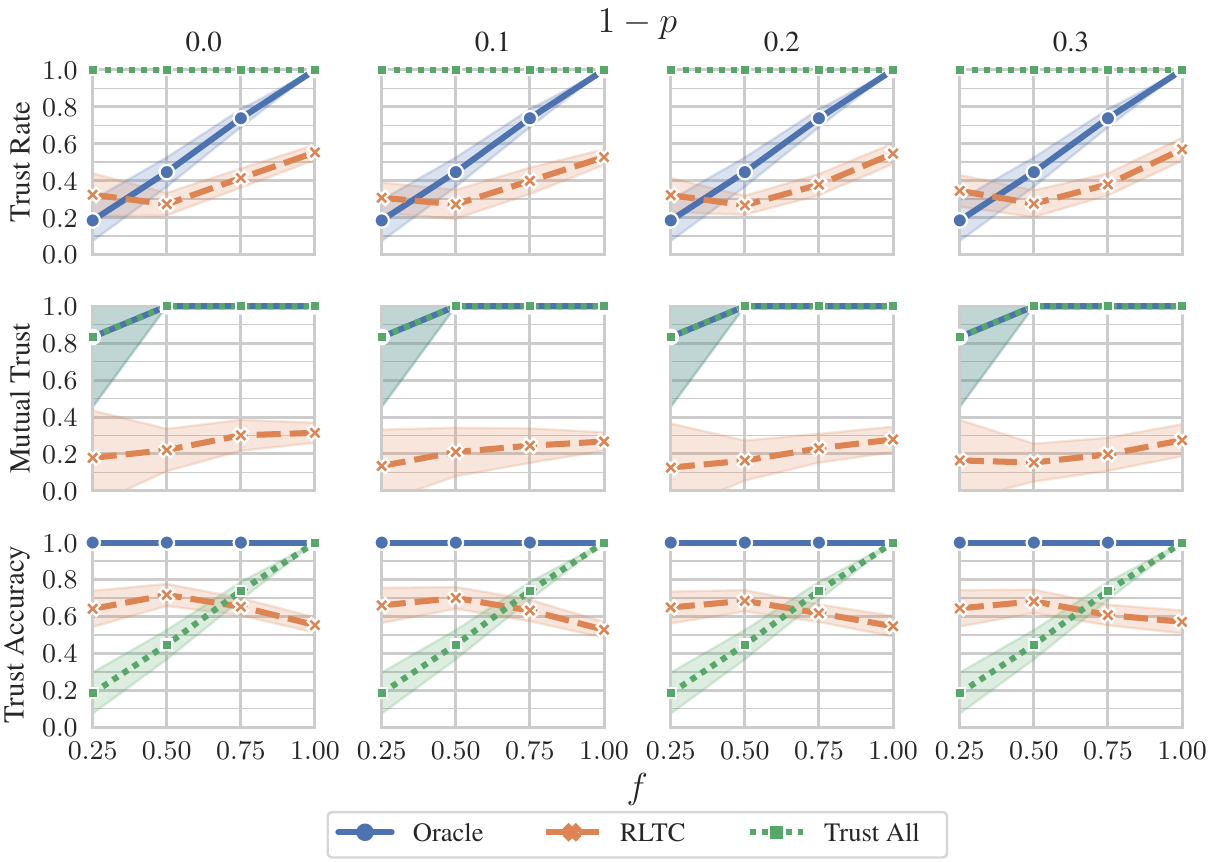}
    \end{center}
    \caption{$f$ against trust metrics for fixed $1-p$ values with 16 agents and \emph{Random} failure model.}
    \label{fig:results-frac-random}
\end{figure}

\section{Experiments}
Our goal is to determine whether the proposed decentralized trust mechanism 
can increase the chances of successful consensus despite the presence of unreliable agents, for either the \emph{Fixed} or \emph{Random} failure model.
To this end, we implement an environment according to our problem specification and perform a series of MARL experiments using the setup described in Section \ref{section:marl-model}.\footnote{The source code for fully reproducing the reported results will be made publicly available in a future version.}

\subsection{Experimental Setup}

\begin{wrapfigure}{r}{0.38\linewidth}
    \vspace{-\baselineskip}
        \centering
        \begin{subfigure}{0.45\linewidth}
            \centering
            \includegraphics[width=\textwidth]{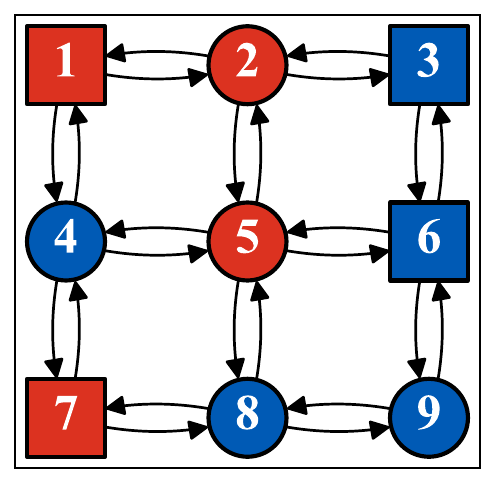}
            \caption{\emph{Trust All}}
        \end{subfigure}
        \hfill
        \begin{subfigure}{0.45\linewidth}
            \centering
            \includegraphics[width=\textwidth]{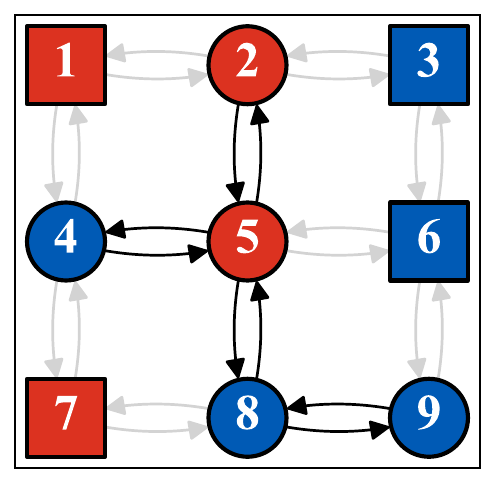}
            \caption{\emph{Oracle}}
        \end{subfigure}
        \caption{
            Scenarios with baseline agents with fixed trust score arrays (circles) and unreliable agents (squares).
        }
        \label{fig:baseline-agents}
    \vspace{-\baselineskip}
\end{wrapfigure}

We run experiments on both $3 \times 3$ and $4 \times 4$ square lattices
($N = 9$ and 16 respectively) and both \emph{Fixed} and \emph{Random} failure models.
We vary two parameters, namely the fraction $f$ of reliable agents in the system and the value initialization noise $1 - p$.
For each failure model, we measure the average consensus success rate and also observe the emergent properties of the learned trust mechanisms.
We repeat the experiments 30 times, each one starting with a unique random seed.
At the start of each repetition, the reliable and unreliable agents are randomly positioned in the communication graph.
Each repetition executes 20,000 training episodes and 2,000 evaluation episodes.
Results are averaged over the random seeds.
The performances of our method \emph{RLTC} (i.e., trust learned by using IQL) are compared with two \emph{fixed} baselines:
\emph{Trust All} where all agents trust each other, and \emph{Oracle} where each reliable agent knows which of its neighbors are reliable a priori and only trusts them (see Figure \ref{fig:baseline-agents} for an example).
These baseline agents are \emph{not} trained using RL and their trust scores are always fixed.
Please refer to the supplementary material for additional experimental details.

Table \ref{table:metrics} outlines the metrics that are recorded in our experiments.
We assess both the consensus success rate and properties of the trust system that emerges. All metrics are computed exclusively over the subset of \emph{reliable} agents. For the \emph{Trust All} baseline, the average trust rate is always 1 since all agents trust each other.
For the \emph{Oracle}, average trust accuracy is 1.
This metric describes how far a learned solution differs from the oracle.
During the experiments, each metric is computed per timestep, then averaged over all timesteps for an episode-specific statistic.

\subsection{Results}

\noindent \textbf{Fixed failure model results.} We first present the experimental results for 16 agents and the \emph{Fixed} failure model.
The observed trends in the 9-agent experiments are similar, thus are moved to the supplementary material.
Figure \ref{fig:success-rates} (top) and Figure \ref{fig:frac-reliable} show the effects of varying the fraction of reliable agents
on the performance metrics for several fixed values of noise.
The values for $f$ and $1-p$ are
$\{0.25, 0.5, 0.75, 1.0\}$ and $\{0, 0.1, 0.2, 0.3\}$ respectively.

We observe that fixing the fraction of reliable agents and varying noise has little effect on the output metrics for \emph{RLTC} agents.
Increasing $1-p$ only decreases the success rate for \emph{Oracle} and \emph{Trust All} agents, which is expected behavior.
For fixed $1-p$, as $f$ increases, the success rate increases as for \emph{Trust All} and \emph{RLTC}; we also note that the success rate for \emph{RLTC} agents is higher than \emph{Trust All}, suggesting that the trust mechanism leads to a statistically significant improvement.
In addition, both the average trust rate and mutual trust increase as the reliable fraction increases.
However, the values are low compared to the oracle,
meaning that the reliable agents do not always trust each other
and often prefer their own values $v_i$ over those of their neighbors when performing updates.
Low mutual trust also indicates the lack of reciprocation between agents.
This is challenging due to our decentralized setup:
each agent updates its trust scores locally and independently of others, hence it is difficult to coordinate trust between pairs of agents.
The trust accuracy for \emph{RLTC} agents is also relatively low, which means that it deviates from the oracle solution.
It shows a decreasing trend as the fraction of reliable agents increases until $f = 0.75$,
but increases between $0.75 \leq f \leq 1$ (refer to the supplementary material for a further discussion about this phenomenon).

\begin{wrapfigure}{r}{0.38\linewidth}
    \vspace{-0.75\baselineskip}
    \centering
    \resizebox{0.99\linewidth}{!}{
        \includegraphics{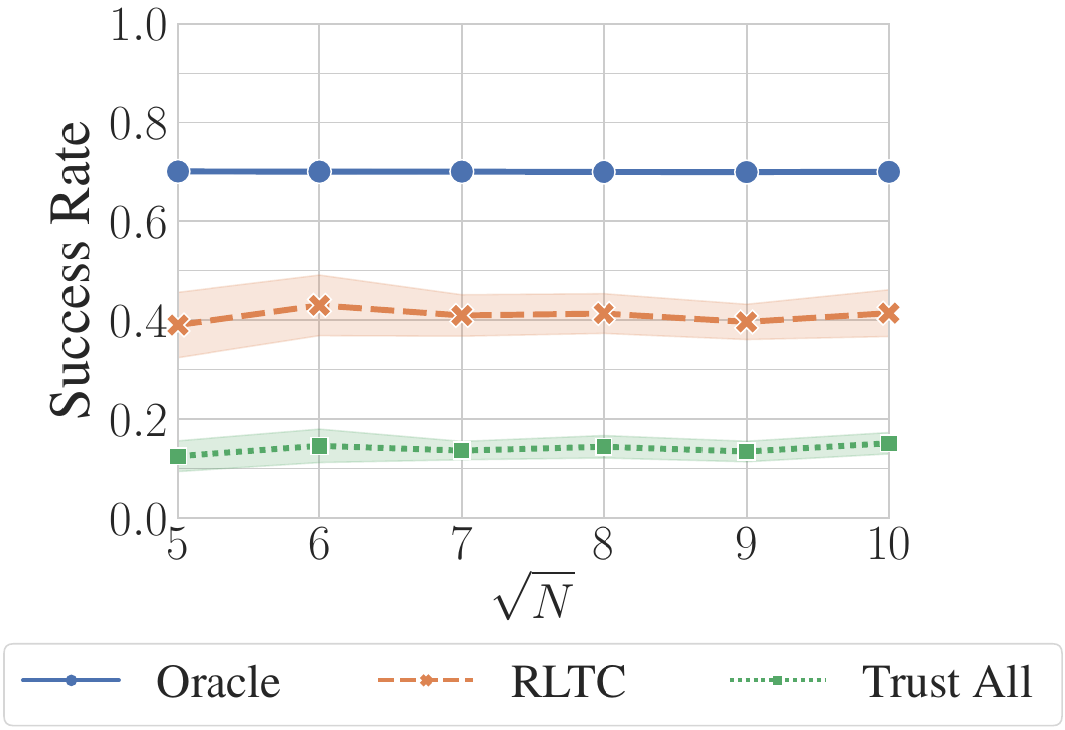}
    }
    \caption{Success rates against the grid dimension $\sqrt{N}$.}
    \label{fig:results-scalability}
\end{wrapfigure}

\noindent \textbf{Random failure model results.}
Similarly, for the \emph{Random} failure model, we present these results for 16 agents in Figure \ref{fig:success-rates} (bottom) and Figure \ref{fig:results-frac-random}.
Notice that the success rates are higher in general because the unreliable agents output the correct value 50\% of the time.
However, \emph{RLTC} exhibits a statistically significant increase in success rate from \emph{Trust All}.
This shows that the functionality of our method is not limited to one type of failure model.

\noindent \textbf{Scalability.}
We also investigate the fundamental dimension of scalability by experimenting with larger communication grid sizes ($5, 6, 7,8,9, 10$) with parameters $f = 0.75$ and $1 - p = 0.3$ as an example. The results shown in Figure \ref{fig:results-scalability} demonstrate that our decentralized approach maintains similar performance as the number of agents increases.

\section{Conclusion}

In this paper, we have investigated the problem of learning consensus in the presence of unreliable agents, a largely neglected yet fundamental issue for the deployment of real-world MARL systems. %
We have presented \emph{Reinforcement Learning-based Trusted Consensus} (RLTC), a decentralized emergent trust mechanism that allows agents to independently learn which neighbors to trust and which to ignore.
Our experiments show that the trust mechanism significantly improves the consensus success rate, indicating that we can deal with unreliable agents effectively while generalizing to different types of failure models and scaling to systems with a greater number of agents.
We have also studied properties of the emerging protocol, showing that it leads to low overall trust rate as well as asymmetric trust between agents.
Our findings highlight that the proposed trust mechanism enables a decentralized protocol to emerge, which can reduce the impact of unreliable agents in a multi-agent system. RLTC has the potential to be used as a modular component for other, more complex tasks that involve agent coordination.

\bibliography{main}

\begin{thebibliography}{46}
\providecommand{\natexlab}[1]{#1}
\providecommand{\url}[1]{\texttt{#1}}
\expandafter\ifx\csname urlstyle\endcsname\relax
  \providecommand{\doi}[1]{doi: #1}\else
  \providecommand{\doi}{doi: \begingroup \urlstyle{rm}\Url}\fi

\bibitem[Abdul-Rahman \& Hailes(1998)Abdul-Rahman and Hailes]{Rahman1998Trust}
Alfarez Abdul-Rahman and Stephen Hailes.
\newblock A distributed trust model.
\newblock In \emph{Proceedings of the 1997 Workshop on New Security Paradigms},
  pp.\  48–60. Association for Computing Machinery, 1998.

\bibitem[Barrat et~al.(2008)Barrat, Barthelemy, and
  Vespignani]{Barrat2009ComplexNetworks}
Alain Barrat, Marc Barthelemy, and Alessandro Vespignani.
\newblock \emph{Dynamical Processes on Complex Networks}.
\newblock Cambridge University Press, 2008.

\bibitem[Blumenkamp \& Prorok(2021)Blumenkamp and
  Prorok]{blumenkamp2021emergence}
Jan Blumenkamp and Amanda Prorok.
\newblock The emergence of adversarial communication in multi-agent
  reinforcement learning.
\newblock In \emph{Proceedings of the 5th Conference on Robot Learning
  (CoRL'21)}, pp.\  1394--1414, 2021.

\bibitem[Bouchacourt \& Baroni(2018)Bouchacourt and Baroni]{Bouchacourt2019}
Diane Bouchacourt and Marco Baroni.
\newblock {How agents see things: On visual representations in an emergent
  language game}.
\newblock In \emph{Proceedings of the 2018 Conference on Empirical Methods in
  Natural Language Processing (EMNLP'18)}, pp.\  981--985. Association for
  Computational Linguistics, 2018.

\bibitem[Bullo(2022)]{Bullo2022Networks}
Francesco Bullo.
\newblock \emph{Lectures on Network Systems}.
\newblock Kindle Direct Publishing, 1.6 edition, 2022.
\newblock URL \url{http://motion.me.ucsb.edu/book-lns}.

\bibitem[Cachin et~al.(2011)Cachin, Guerraoui, and
  Rodrigues]{Cachin2011Reliable}
Christian Cachin, Rachid Guerraoui, and Luis Rodrigues.
\newblock \emph{Introduction to Reliable and Secure Distributed Programming}.
\newblock Springer, 2011.

\bibitem[Castelfranchi \& Falcone(1998)Castelfranchi and
  Falcone]{Castelfranchi1998Trust}
C.~Castelfranchi and R.~Falcone.
\newblock Principles of trust for {{MAS}}: cognitive anatomy, social
  importance, and quantification.
\newblock In \emph{Proceedings of the IEEE International Conference on Multi
  Agent Systems}, pp.\  72--79, 1998.

\bibitem[Christianos et~al.(2020)Christianos, Sch\"{a}fer, and
  Albrecht]{Christianos2020shared}
Filippos Christianos, Lukas Sch\"{a}fer, and Stefano Albrecht.
\newblock Shared experience actor-critic for multi-agent reinforcement
  learning.
\newblock In \emph{Advances in Neural Information Processing Systems},
  volume~33, 2020.

\bibitem[Cohen et~al.(2019)Cohen, Schaekermann, Liu, and
  Cormier]{Cohen2019Trust}
Robin Cohen, Mike Schaekermann, Sihao Liu, and Michael Cormier.
\newblock Trusted {{AI}} and the contribution of trust modeling in multiagent
  systems.
\newblock In \emph{Proceedings of the 18th International Conference on
  Autonomous Agents and Multi-Agent Systems (AAMAS'19)}, pp.\  1644–1648,
  2019.

\bibitem[Coulouris et~al.(2012)Coulouris, Dollimore, Kindberg, and
  Blair]{Coulouris2012}
George Coulouris, Jean Dollimore, Tim Kindberg, and Gordon Blair.
\newblock \emph{Distributed Systems: Concepts and Design}.
\newblock Pearson Education, 2012.

\bibitem[Dafoe et~al.(2020)Dafoe, Hughes, Bachrach, Collins, McKee, Leibo,
  Larson, and Graepel]{Dafoe2020OpenCoop}
Allan Dafoe, Edward Hughes, Yoram Bachrach, Tantum Collins, Kevin~R McKee,
  Joel~Z Leibo, Kate Larson, and Thore Graepel.
\newblock {Open Problems in Cooperative AI}.
\newblock \emph{arXiv preprint arXiv:2012.08630}, 2020.

\bibitem[Das et~al.(2019)Das, Gervet, Romoff, Batra, Parikh, Rabbat, and
  Pineau]{Das2019Tarmac}
Abhishek Das, Th{\'e}ophile Gervet, Joshua Romoff, Dhruv Batra, Devi Parikh,
  Mike Rabbat, and Joelle Pineau.
\newblock {TarMAC}: Targeted multi-agent communication.
\newblock In \emph{Proceedings of the 36th International Conference on Machine
  Learning (ICML'19)}, pp.\  1538--1546, 2019.

\bibitem[DeGroot(1974)]{DeGroot1974Social}
Morris~H. DeGroot.
\newblock Reaching a consensus.
\newblock \emph{Journal of the American Statistical Association}, 69\penalty0
  (345), 1974.

\bibitem[Fax \& Murray(2004)Fax and Murray]{Fax2004Vehicles}
J.A. Fax and R.M. Murray.
\newblock Information flow and cooperative control of vehicle formations.
\newblock \emph{IEEE Transactions on Automatic Control}, 49\penalty0 (9), 2004.

\bibitem[Foerster et~al.(2016)Foerster, Assael, de~Freitas, and
  Whiteson]{Foerster2016DIAL}
Jakob Foerster, Ioannis~Alexandros Assael, Nando de~Freitas, and Shimon
  Whiteson.
\newblock Learning to communicate with deep multi-agent reinforcement learning.
\newblock In \emph{Advances in Neural Information Processing Systems},
  volume~29, 2016.

\bibitem[Graesser et~al.(2019)Graesser, Cho, and Kiela]{Graesser2019}
Laura~Harding Graesser, Kyunghyun Cho, and Douwe Kiela.
\newblock {Emergent Linguistic Phenomena in Multi-Agent Communication Games}.
\newblock In \emph{Proceedings of the 2019 Conference on Empirical Methods in
  Natural Language Processing (EMNLP'19)}, pp.\  3700--3710. Association for
  Computational Linguistics, 2019.

\bibitem[Haahr(1998--2018)]{RandomOrg}
Mads Haahr.
\newblock {RANDOM.ORG:} true random number service.
\newblock \url{https://www.random.org}, 1998--2018.
\newblock Accessed: 2024-03-06.

\bibitem[Hagberg et~al.(2008)Hagberg, Swart, and
  S.~Chult]{hagberg2008exploring}
Aric Hagberg, Pieter Swart, and Daniel S.~Chult.
\newblock Exploring network structure, dynamics, and function using networkx.
\newblock In \emph{SciPy}, 2008.

\bibitem[Harris et~al.(2020)Harris, Millman, van~der Walt, Gommers, Virtanen,
  Cournapeau, Wieser, Taylor, Berg, Smith, et~al.]{harris2020array}
Charles~R Harris, K~Jarrod Millman, St{\'e}fan~J van~der Walt, Ralf Gommers,
  Pauli Virtanen, David Cournapeau, Eric Wieser, Julian Taylor, Sebastian Berg,
  Nathaniel~J Smith, et~al.
\newblock {Array programming with NumPy}.
\newblock \emph{Nature}, 585\penalty0 (7825):\penalty0 357--362, 2020.

\bibitem[Hunter(2007)]{hunter2007mpl}
J.~D. Hunter.
\newblock {Matplotlib: A 2D graphics environment}.
\newblock \emph{Computing in Science \& Engineering}, 9\penalty0 (3):\penalty0
  90--95, 2007.

\bibitem[Krapivsky \& Redner(2003)Krapivsky and Redner]{Krapivsky2003Majority}
Paul~L Krapivsky and Sidney Redner.
\newblock Dynamics of majority rule in two-state interacting spin systems.
\newblock \emph{Physical Review Letters}, 90\penalty0 (23):\penalty0 238701,
  2003.

\bibitem[Lamport(1998)]{Lamport1998Paxos}
Leslie Lamport.
\newblock The part-time parliament.
\newblock \emph{ACM Transactions on Computer Systems}, 16\penalty0
  (2):\penalty0 133–169, 1998.

\bibitem[Lamport et~al.(1982)Lamport, Shostak, and Pease]{LamportByzantine}
Leslie Lamport, Robert Shostak, and Marshall Pease.
\newblock The {B}yzantine generals problem.
\newblock \emph{ACM Transactions on Programming Languages and Systems},
  4\penalty0 (3):\penalty0 382–401, 1982.

\bibitem[Lazaridou et~al.(2018)Lazaridou, Hermann, Tuyls, and
  Clark]{Lazaridou2018}
Angeliki Lazaridou, Karl~Moritz Hermann, Karl Tuyls, and Stephen Clark.
\newblock {Emergence of Linguistic Communication from Referential Games with
  Symbolic and Pixel Input}.
\newblock In \emph{Proceedings of the 6th International Conference on Learning
  Representations (ICLR'18)}, 2018.

\bibitem[Lianza \& Snook(2020)Lianza and Snook]{Cloudflare2020Byzantine}
Tom Lianza and Chris Snook.
\newblock A {B}yzantine failure in the real world - the cloudflare blog, Nov
  2020.
\newblock URL
  \url{https://blog.cloudflare.com/a-byzantine-failure-in-the-real-world/}.
\newblock Last accessed: 06 March 2024.

\bibitem[Lowe et~al.(2017)Lowe, Wu, Tamar, Harb, Abbeel, and
  Mordatch]{Lowe2017multi}
Ryan Lowe, Yi~Wu, Aviv Tamar, Jean Harb, Pieter Abbeel, and Igor Mordatch.
\newblock Multi-agent actor-critic for mixed cooperative-competitive
  environments.
\newblock In \emph{Advances in Neural Information Processing Systems},
  volume~30, 2017.

\bibitem[McKinney et~al.(2011)]{mckinney2011pandas}
Wes McKinney et~al.
\newblock pandas: a foundational {Python} library for data analysis and
  statistics.
\newblock \emph{Python for High Performance and Scientific Computing},
  14\penalty0 (9):\penalty0 1--9, 2011.

\bibitem[Mordatch \& Abbeel(2017)Mordatch and Abbeel]{Mordatch2017}
Igor Mordatch and Pieter Abbeel.
\newblock {Emergence of Grounded Compositional Language in Multi-Agent
  Populations}.
\newblock \emph{Proceedings of the 31st AAAI Conference on Artificial
  Intelligence (AAAI'17)}, pp.\  1495--1502, 2017.

\bibitem[Olfati-Saber \& Shamma(2005)Olfati-Saber and
  Shamma]{Olfati2005Sensors}
R.~Olfati-Saber and J.S. Shamma.
\newblock Consensus filters for sensor networks and distributed sensor fusion.
\newblock In \emph{Proceedings of the 44th IEEE Conference on Decision and
  Control (CDC'05)}, 2005.

\bibitem[Olfati-Saber et~al.(2007)Olfati-Saber, Fax, and
  Murray]{Olfati2007NetworkedConsensus}
Reza Olfati-Saber, J.~Alex Fax, and Richard~M. Murray.
\newblock Consensus and cooperation in networked multi-agent systems.
\newblock \emph{Proceedings of the IEEE}, 95\penalty0 (1):\penalty0 215--233,
  Jan 2007.

\bibitem[Ongaro \& Ousterhout(2014)Ongaro and Ousterhout]{Ongaro2014Raft}
Diego Ongaro and John Ousterhout.
\newblock In search of an understandable consensus algorithm.
\newblock In \emph{2014 {USENIX} Annual Technical Conference}, pp.\  305--319.
  {USENIX} Association, 2014.

\bibitem[Pease et~al.(1980)Pease, Shostak, and Lamport]{PSL1980AgreeFaults}
Marshall Pease, Robert Shostak, and Leslie Lamport.
\newblock Reaching agreement in the presence of faults.
\newblock \emph{Journal of the ACM}, 27\penalty0 (2):\penalty0 228–234, 1980.

\bibitem[Pinto et~al.(2017)Pinto, Davidson, Sukthankar, and
  Gupta]{pinto2017robust}
Lerrel Pinto, James Davidson, Rahul Sukthankar, and Abhinav Gupta.
\newblock Robust adversarial reinforcement learning.
\newblock In \emph{Proceedings of the 34th International Conference on Machine
  Learning (ICML'17)}, pp.\  2817--2826, 2017.

\bibitem[Ramchurn et~al.(2004)Ramchurn, Huynh, and Jennings]{Ramchurn2004Trust}
Sarvapali~D. Ramchurn, Dong Huynh, and Nicholas~R. Jennings.
\newblock Trust in multi-agent systems.
\newblock \emph{Knowledge Engineering Review}, 19\penalty0 (1):\penalty0
  1–25, 2004.

\bibitem[Rangwala \& Williams(2020)Rangwala and Williams]{RangwalaWilliams2020}
Murtaza Rangwala and Ryan Williams.
\newblock Learning multi-agent communication through structured attentive
  reasoning.
\newblock In \emph{Advances in Neural Information Processing Systems},
  volume~33, 2020.

\bibitem[Ren et~al.(2005)Ren, Beard, and Atkins]{Ren2005Survey}
Wei Ren, R.W. Beard, and E.M. Atkins.
\newblock A survey of consensus problems in multi-agent coordination.
\newblock In \emph{Proceedings of the 2005 American Control Conference
  (ACC'05)}, pp.\  1859--1864 vol. 3, 2005.

\bibitem[Sabater \& Sierra(2005)Sabater and Sierra]{Sabater2005TrustReview}
Jordi Sabater and Carles Sierra.
\newblock Review on computational trust and reputation models.
\newblock \emph{Artificial Intelligence Review}, 24\penalty0 (1):\penalty0
  33--60, 2005.

\bibitem[Schroeder \& Gibson(2010)Schroeder and Gibson]{Schroeder2010FailHPC}
Bianca Schroeder and Garth~A. Gibson.
\newblock A large-scale study of failures in high-performance computing
  systems.
\newblock \emph{IEEE Transactions on Dependable and Secure Computing},
  7\penalty0 (4):\penalty0 337--350, 2010.

\bibitem[Sukhbaatar et~al.(2016)Sukhbaatar, Szlam, and Fergus]{Sukhbaatar2016}
Sainbayar Sukhbaatar, Arthur Szlam, and Rob Fergus.
\newblock Learning multiagent communication with backpropagation.
\newblock In \emph{Advances in Neural Information Processing Systems},
  volume~29, 2016.

\bibitem[Sun et~al.(2023)Sun, Zheng, Hassanzadeh, Liang, Feizi, Ganesh, and
  Huang]{sun2023certifiably}
Yanchao Sun, Ruijie Zheng, Parisa Hassanzadeh, Yongyuan Liang, Soheil Feizi,
  Sumitra Ganesh, and Furong Huang.
\newblock Certifiably robust policy learning against adversarial multi-agent
  communication.
\newblock In \emph{Proceedings of the 11th International Conference on Learning
  Representations (ICLR'23)}, 2023.

\bibitem[Tan(1993)]{Tan1993IndepQ}
Ming Tan.
\newblock Multi-agent reinforcement learning: Independent vs. cooperative
  agents.
\newblock In \emph{Proceedings of the 10th International Conference on Machine
  Learning (ICML'93)}, pp.\  330--337, 1993.

\bibitem[Waskom(2021)]{waskom2021seaborn}
Michael~L. Waskom.
\newblock Seaborn: statistical data visualization.
\newblock \emph{Journal of Open Source Software}, 6\penalty0 (60):\penalty0
  3021, 2021.

\bibitem[Watkins \& Dayan(1992)Watkins and Dayan]{Watkins1992}
Christopher J. C.~H. Watkins and Peter Dayan.
\newblock {Q}-learning.
\newblock \emph{Machine Learning}, 8\penalty0 (3):\penalty0 279--292, 1992.

\bibitem[Xue et~al.(2022)Xue, Qiu, An, Rabinovich, Obraztsova, and
  Yeo]{xue2022mis}
Wanqi Xue, Wei Qiu, Bo~An, Zinovi Rabinovich, Svetlana Obraztsova, and
  Chai~Kiat Yeo.
\newblock Mis-spoke or mis-lead: Achieving robustness in multi-agent
  communicative reinforcement learning.
\newblock In \emph{Proceedings of the 21st International Conference on
  Autonomous Agents and Multi-Agent Systems (AAMAS'22)}, 2022.

\bibitem[Zhang et~al.(2020{\natexlab{a}})Zhang, Chen, Xiao, Li, Liu, Boning,
  and Hsieh]{zhang2020robust}
Huan Zhang, Hongge Chen, Chaowei Xiao, Bo~Li, Mingyan Liu, Duane Boning, and
  Cho-Jui Hsieh.
\newblock Robust deep reinforcement learning against adversarial perturbations
  on state observations.
\newblock In \emph{Advances in Neural Information Processing Systems},
  volume~33, 2020{\natexlab{a}}.

\bibitem[Zhang et~al.(2020{\natexlab{b}})Zhang, Zhang, and Lin]{Zhang2020I2C}
Sai~Qian Zhang, Qi~Zhang, and Jieyu Lin.
\newblock Succinct and robust multi-agent communication with temporal message
  control.
\newblock In \emph{Advances in Neural Information Processing Systems},
  volume~33, 2020{\natexlab{b}}.

\end{thebibliography}
\bibliographystyle{rlc}

\appendix

\section{Experimental Setting}

\noindent \textbf{Implementation. } In a future version, the source code will be made publicly available, which will enable the reproduction of all results presented in the paper, including tables and figures. Our implementation uses public software libraries
\citep{hagberg2008exploring,harris2020array,mckinney2011pandas,hunter2007mpl,waskom2021seaborn}.

\noindent \textbf{Platform specifications. }
Experiments are run on Linux (CentOS 7.9.2009) using Python 3.10.3.
The platform uses Intel(R) Xeon(R) CPU E5-2637 v4 @ 3.50GHz with 16 cores and 60 GB RAM.
Training occurs purely on CPU.
The full set of experiments takes approximately 12 hours to run.

\noindent \textbf{Configuration.}
\cref{table:rl-hyperparameters} lists the hyperparameters used in all experiments.
Multiplicative epsilon decay is used, i.e. $\varepsilon_{t+1} = r^t \varepsilon_0$ at each timestep,
where $r$ is a constant decay factor.
20,000 training episodes and 2,000 evaluation episodes are used,
each episode with length 30.
Results are averaged over 30 random seeds,
which were generated using \citet{RandomOrg} in the source code.
\cref{table:variable-ranges} reiterates the values that were used in experiments
from the main paper for convenience (excluding the scalability experiment, which uses more values of $N$.)

\begin{table}[ht]
    \caption{RL hyperparameters.}
    \label{table:rl-hyperparameters}
    \centering
    \begin{tabular}{p{0.5\linewidth} p{0.2\linewidth}}
        \toprule
        Name & Value \\
        \midrule
        $Q$-learning step size $\alpha$ & 0.03 \\
        Discount factor $\gamma$ & 0.999 \\
        Initial exploration probability $\varepsilon_0$ & 0.3 \\
        Exploration decay factor $r$ & 0.9996 \\

        \bottomrule
    \end{tabular}
\end{table}

\begin{table}[ht]
    \caption{Variable ranges.}
    \label{table:variable-ranges}
    \centering
        \begin{tabular}{p{0.55\linewidth} p{0.3\linewidth}}
            \toprule
            Name & Values \\
            \midrule
            Number of agents $N$ & 9, 16\\
            Fraction of reliable agents $f$ & 0.25, 0.5, 0.75, 1.0 \\
            Local value initialization noise $1 - p$ & 0, 0.1, 0.2, 0.3 \\
            \bottomrule
        \end{tabular}
\end{table}

\noindent \textbf{Hyperparameter selection.}
The RL hyperparameters are selected using a grid search.
See \cref{table:rl-gridsearch} for the combinations used.
We discover that only the discount factor $\gamma$ and epsilon decay factor $r$
have any noticeable impact on the average reward:
using $\gamma = 0.999$ and $r = 0.9996$ converges to the highest value.
$\alpha = 0.03$ leads to less noisy rewards over time, 
and a larger $\varepsilon_0$ allows for more initial exploration.

\begin{table}[ht]
    \caption{RL hyperparameter grid search.}
    \label{table:rl-gridsearch}
    \centering
    \begin{tabular}{p{0.15\linewidth} p{0.3\linewidth}}
        \toprule
        Name & Values \\
        \midrule
        $\alpha$ & 0.03, 0.01, 0.1 \\
        $\gamma$ & 0.999, 0.95 \\
        $\varepsilon_0$ & 0.1, 0.3 \\
        $r$ & 0.9996, 1.0 \\
        \bottomrule
    \end{tabular}
\end{table}

\section{Additional Results and Figures}

\noindent \textbf{Results for 9 agents, fixed failure model.}
\cref{fig:noise-9} and \cref{fig:frac-9} show the experimental results for 9 agents (fixed failure model)
that were omitted in the main text for brevity,
mainly due to similar trends being exhibited.

\begin{figure*}[pt]
    \begin{center}
        \input{figures/noise_9_agents.pgf}
    \end{center}
    \caption{Noise against metric for fixed reliable fractions (9 agents, fixed failure model).}
    \label{fig:noise-9}
\end{figure*}

\begin{figure*}[pt]
    \begin{center}
        \input{figures/frac_reliable_9_agents.pgf}
    \end{center}
    \caption{Reliable fraction against metric for fixed noise values (9 agents, fixed failure model).}
    \label{fig:frac-9}
\end{figure*}

\noindent \textbf{Plots for 16 agents, fixed failure model, but with constant $f$ per column.}
\cref{fig:noise-16} shows the same results for 16 agents as seen in the main text,
but with $1-p$ on the $x$-axes and fixed $f$ per column.

\begin{figure*}[pt]
    \begin{center}
        \input{figures/noise_16_agents.pgf}
    \end{center}
    \caption{Noise against metric for fixed reliable fractions (16 agents, fixed failure model).}
    \label{fig:noise-16}
\end{figure*}

\noindent \textbf{Plots for 16 agents, randomized failure model, but with constant $f$ per column.}
\cref{fig:results-noise-random} shows the same results for the random failure model and 16 agents as seen in the main text,
but with $1-p$ on the $x$-axes and fixed $f$ per column.

\begin{figure*}[pt]
    \begin{center}
            \input{figures/noise_16_agents_random.pgf}
    \end{center}
    \caption{$1-p$ against metric for fixed $f$ values (16 agents, randomized failure model).}
    \label{fig:results-noise-random}
\end{figure*}

\noindent \textbf{Investigating trust accuracies for $0.75 \leq f \leq 1.0$.}
In the main paper, we notice that the trust accuracy of trained agents
decreases when the fractions of reliable agents is between 0 and 0.75,
but increases in the 0.75 to 1.0 interval.
To investigate this inconsistency, we run finer-grained experiments
for values 0.75, 0.8, 0.85, 0.9, 0.95, 1.0.
The results are illustrated in \cref{fig:trust-accuracy-debug}.
We notice a general smooth increase for fixed values of noise,
which is an interesting emergent property of the system.

\begin{figure*}[ht]
    \begin{center}
        \resizebox*{\linewidth}{!}{
          \input{figures/trust_accuracy_debug.pgf}
        }
    \end{center}
    \caption{Trust accuracy for $0.75 \leq f \leq 1.0$, 16 agents.}
    \label{fig:trust-accuracy-debug}
\end{figure*}

\end{document}